\documentclass[aps,prl,twocolumn,superscriptaddress,showpacs,floatfix]{revtex4-1}
\usepackage{graphicx,amsmath,amssymb,xspace,epsfig,float,amsthm,latexsym,subfigure}

\newcommand{\B}[1]{\mathbf{#1}}

\begin{document}

\title{Dynamical spin-flip susceptibility for a strongly interacting ultracold Fermi gas}

\author{Matteo Sandri}
\affiliation{International School for Advanced Studies (SISSA) - Via Bonomea 265,  I-34136 Trieste, Italy}
\affiliation{Dipartimento di Fisica “Galileo Galilei” and CNISM, Universit\`a di Padova, Via Marzolo 8, 35122 Padova, Italy}

\author{Anna Minguzzi}
\affiliation{Universit\'e Grenoble I and CNRS, Laboratoire de Physique et
  Mod\'elisation, des Milieux Condens\'es UMR 5493, Maison des Magist\`eres, B.P.\ 166, 38042 Grenoble,
  France}

\author{Flavio Toigo}
\affiliation{Dipartimento di Fisica “Galileo Galilei” and CNISM, Universit\`a di Padova, Via Marzolo 8, 35122 Padova, Italy}

\date{\today}

\begin{abstract}
The Stoner model predicts that a two-component Fermi gas at increasing repulsive interactions undergoes a ferromagnetic transition. Using the random-phase approximation we study the dynamical properties of the interacting Fermi gas. For an atomic Fermi gas under harmonic confinement we show that the transverse (spin-flip) dynamical susceptibility displays a clear signature of the ferromagnetic phase in a magnon peak emerging from the Stoner particle-hole continuum. The dynamical spin susceptibilities could be experimentally explored via  spin-dependent Bragg spectroscopy.
\end{abstract}


\maketitle

{\it Introduction} 
Experimental advances of trapping and cooling ultracold Fermi
 gases allow the exploration of strongly interacting quantum degenerate  
two-component gases, realized  by populating two  hyperfine atomic levels.  The strength and  the sign of the intercomponent interactions can be tuned 
by the mechanism of Feschbach resonances. The case of attractive 
Fermi-Fermi interactions has allowed the exploration of the crossover from BCS-like  pairing to Bose-Einstein condensation (BEC)  of tightly bound pairs \cite{Zwerger:2}. On the repulsive side of the Feschbach resonance,  the metastable branch corresponding to a two-component Fermi gas with repulsive interactions has received more recent experimental attention \cite{Ketterle:1,Sommer11} as a possible realization of the Stoner model of itinerant ferromagnetism.
 The mean-field Stoner model  \cite{Stoner38} predicts that a repulsive Fermi gas undergoes a transition to a ferromagnetic state for sufficiently large interaction strength $k_F a > \pi/2$, where $a$ is the $s$-wave scattering length and $k_F=(3 \pi^2 n )^{1/3}$ is the Fermi wavevector for a uniform gas with total density $n$. Beyond mean field corrections, the  inclusion of quadratic fluctuations decreases this value to  $k_Fa \approx 1.05$  \cite{Conduit:1,Duine:1}, and Quantum Monte Carlo  simulations predict the transition to occur at $k_F \,a \approx 0.82$ \cite{Pilati:1,Chang11}. While these calculations assume a uniform ferromagnetic phase, the possibility of inhomogeneous intermediate phases has been considered \cite{Conduit09}.

In this work we analyze the dynamical properties of a two-component Fermi  gas with repulsive interactions. The approach to the ferromagnetic transition strongly affects the dynamical properties of the system, displaying a softening of the out-of phase spin-dipole modes   \cite{Recati:1} and enhancement of the  spin drag coefficient \cite{Duine:2}. 
 A dynamical instability of a spin spiral excitation has also been predicted \cite{Conduit10}. Above the ferromagnetic transition, magnon collective modes are predicted to propagate undamped at zero temperature \cite{Callaway:1}.

Spectroscopy is a powerful tool to address many-body systems. The study of the  dynamic structure factor for density and spin has been suggested as a probe of pairing and superfluidity of a Fermi gas \cite{Minguzzi01,Zwerger04}, and  Bragg scattering experiments for a strongly interacting Fermi gas in the BCS-BEC crossover have been performed \cite{Veeravalli08}.

We focus here on the spin-susceptibility spectra as a tool to characterize the ferromagnetic phase of a two component Fermi gas. Such spectra could be accessed experimentally via spin-dependent Bragg spectroscopy.

{\it Dynamical susceptibilities in the random phase approximation}
We consider a two-component Fermi gas $|\uparrow\rangle$, $|\downarrow\rangle$, at zero temperature, with repulsive intercomponent interactions described by a contact potential of strength $g=4 \pi \hbar^2a /m$ in terms of the scattering length $a>0$, and the atomic mass $m$. The gas is subjected to  an external confining potential $V_{ext}(\mathbf{r})$. The model 
Hamiltonian of the system reads
\begin{eqnarray}
\label{eq:hamiltoniana}
\hat{H} = \int d\mathbf{r} \hat{\psi}^{\dagger}_{\sigma}(\mathbf{r}) \left[ -\frac{\hbar^2}{2m}\nabla^2 + V_{ext}(\mathbf{r}) \right] \hat{\psi}_{\sigma}(\mathbf{r}) \nonumber \\
+ g\int d\mathbf{r} \hat{\psi}^{\dagger}_{\uparrow}(\mathbf{r}) \hat{\psi}^{\dagger}_{\downarrow}(\mathbf{r}) \hat{\psi}_{\downarrow}(\mathbf{r}) \hat{\psi}_{\uparrow}(\mathbf{r}),
\end{eqnarray}
where the fermionic field operators  $ \hat{\psi}_{\downarrow}(\mathbf{r})$,  $ \hat{\psi}_{\uparrow}(\mathbf{r})$ satisfy the usual anticommutation relations $\{ \hat{\psi}_{\sigma}(\B{r}), \hat{\psi}^{\dagger}_{\sigma'}(\B{r}) \} = \delta_{\sigma.\sigma'}\delta(\B{r}-\B{r}')$. The corresponding spin densities are $\sigma_\alpha(\B{r})=\Psi^\dagger(\B{r})\sigma_\alpha\Psi(\B{r})$ in terms of the spinor $\Psi(\B{r}) = 
\left (
\begin{array}{c}
\psi_\uparrow(\B{r})\\
\psi_\downarrow(\B{r})
\end{array}
\right )
$; $\sigma_\alpha$  are the Pauli matrices with $\alpha=0,x,y,z$. For the transverse   (i.e. $x,y$) components it is useful to introduce  the combinations 
$\sigma_\pm=(\sigma_x\pm i \sigma_y)/2$ because they yield the spin-flip amplitudes
\begin{equation}
{\sigma}_+(\B{r}) = \hat{\psi}_\uparrow^\dagger(\B{r})\hat{\psi}_\downarrow(\B{r}) ; \qquad 
{\sigma}_-(\B{r}) = \hat{\psi}_\downarrow^\dagger(\B{r})\hat{\psi}_\uparrow(\B{r}) .
\end{equation}

In linear response theory, corresponding to the regime of weak external drive, 
the dynamical spin susceptibilities are given by the retarded Green's functions
\begin{equation}
\chi_{\alpha\beta}(\B{r},\B{r}',t,t') = -i\theta(t-t') \langle [\sigma_\alpha(\B{r},t), \sigma_\beta(\B{r}',t')] \rangle,
\end{equation}
where the quantum averages $\langle ... \rangle$ are performed on the unperturbed state of the system. Such dynamical correlations functions embed the effect of interactions, and require a many-body calculation. We consider here their expression as obtained by the Random-Phase approximation (RPA) \cite{Kubo:1,Englert:1}, corresponding to time-dependent Hartree-Fock approximation. This analytical approximation is known to satisfy sum rules and well describe the weak-coupling regime. It also accounts for the ferromagnetic transition at the mean-field (Stoner) level, hence, although not quantitatively correct it is expected to qualitatively describe the dynamical properties of the interacting Fermi gas.  

According to the RPA, for a uniform Fermi gas the Fourier transform of spin-flip susceptibility with respect to both spatial and time relative variables  is given by
\begin{equation}
\label{eq:chirpa}
\chi_{+-}(\omega,\B{q}) =\frac{\Pi_{+-}}{1+g\Pi_{+-}}.
\end{equation}
A similar result holds for the total density $\chi_{00}$ and longitudinal-spin $\chi_{zz}$ responses, 
\begin{equation}
\label{eq:chinn}
\chi_{00}(\omega,\B{q}) =\frac{\Pi_{++}+\Pi_{--}+ 2 g\Pi_{++}\Pi_{--} }{1-g^2\Pi_{++}\Pi_{--}},
\end{equation}
\begin{equation}
\label{eq:chizz}
\chi_{zz}(\omega,\B{q}) =\frac{\Pi_{++}+\Pi_{--}- 2 g\Pi_{++}\Pi_{--} }{1-g^2\Pi_{++}\Pi_{--}},
\end{equation}
where the Lindhard functions $\Pi_{\sigma\sigma'}(\omega,\B{q})$ are given by 
\begin{equation}\label{eq:pol_+-}
\Pi_{\sigma\sigma'}(\omega,\B{q}) = \frac{1}{V} \sum_{\B{k}} \frac{ f_\sigma(\xi_{\B{k}})-f_{\sigma'}(\xi_{\B{k+q}})}{\hbar \omega -\xi_{\B{k+q}}^{\sigma'} + \xi_{\B{k}}^{\sigma} + i \eta}
\end{equation}
with $\sigma,\sigma'=\pm $ indicating the spin polarization along the axis of spontaneous magnetization,  $f_\sigma(\xi)=(e^{\beta \xi_\sigma}+1)^{-1}$, $\beta=1/k_BT$ and $\xi_{\B{k}}^\sigma = \epsilon_{\B{k}} - \mu + g n_{-\sigma}$. We have used the symbol $\pm$ to indicate the two magnetization components since  for long transverse decoherence times and in absence of population imbalance the direction of magnetization of the ferromagnet is not necessarily the quantization  direction of the  total angular momentum $\vec F$   of the atoms  \cite{Conduit:1},  used to define the two-component Fermi gas.

Below the ferromagnetic transition, ie in absence of population imbalance we have $\Pi_{+-}=\Pi_{++}=\Pi_{--}\equiv\Pi$, the well-known Lindhard function for noninteracting fermions \cite{Pines}, and $\chi_{00 (zz)}=2 \Pi/(1 \mp g \Pi)$. For repulsive interactions, characterized by  $g>0$, these expressions predict a well-defined propagating collective mode for the density channel (the zero sound) and a damped mode in the spin channel.

The validity of the RPA can be inferred by comparison with the Landau theory of Fermi liquids. The latter allows to calculate the  density (and spin) susceptibility $\chi_{00 (zz)}$  at small energy and long-wavelength below the ferromagnetic transition. The obtained susceptibilities  \cite{Stringari:3} have  the same structure as Eqs.(\ref{eq:chinn}) and (\ref{eq:chizz}), with the  interaction strength $g$ replaced by  suitable Landau parameters $F_0^{\rho(z)}$.

{\it Spin-flip susceptibilities for a uniform Fermi gas above the ferromagnetic transition}

\begin{figure}
\includegraphics[width=0.7\linewidth]{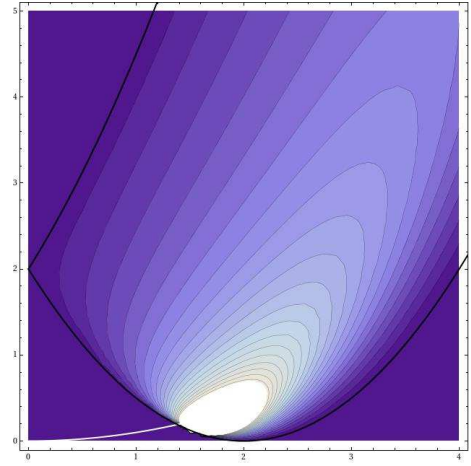}%
\caption{\label{fig1} (Color online) Imaginary part of the spin-flip susceptibility Im$\chi_{+-}(q,\omega)$ for a uniform Fermi gas in the $(\omega,q)$ plane ($q$ in units of $\bar k_F$, $\omega$ in units of $\hbar \bar k_F^2/m$) for the fully polarized case $n_+=n$. The magnon dispersion is indicated by the solid white line.}
\end{figure}

Above the ferromagnetic transition a window opens in the $(\omega,\B{q})$ plane  at small frequencies and wavevectors, below the Stoner particle-hole continuum, allowing for the propagation of spin-flip collective modes (magnons), corresponding to the Goldstone mode of the ferromagnetic phase.  The magnons have a quadratic dispersion $\Omega_q$ which is obtained from the poles of Eq.(\ref{eq:chirpa}), ie $1+g \Pi_{+-}(\Omega_q,\B{q})=0$, yielding  at second order in $q$
\begin{equation}
\hbar \Omega_q = \frac{g}{\Delta} \frac{q^2 \hbar^2}{2m} \left ( (n_++n_-) - \frac{4}{5\Delta} \frac{\hbar^2}{2m} (n_+{k_F^+}^2 - n_-{k_F^-}^2) \right ),
\end{equation}
with $\Delta = g (n_+ - n_-)$, recovering  \cite{Callaway:1} in the fully polarized case, and  $k_F^{\pm} = (6\pi^2 n_\pm)^{1/3}$.

The contribution of the particle-hole continuum is obtained from  the study of  the Lindhard function $\Pi_{+-}$, for which we obtain an analytic expression.
For simplicity of notations we rescale  $|\B{q}|$ in units of $\bar k_F=(k_F^++k_F^-)/2$,  and set $\nu=m\omega/\hbar \bar k_F^2$, $\delta k=(k_F^+-k_F^-)/2\bar k_F$, $d=m g (n_+-n_-)/\hbar \bar k_F^2$. The Lindhard function at zero temperature is nonzero only in a restricted part of the $(\nu,q)$ plane, delimited by the curves $\nu_+^{(1,2)}=q^2/2+d\pm q(1+\delta k)$, $\nu_-^{(1,2)}=-q^2/2+d\mp q(1-\delta k)$.
 In detail,  by setting  $A_\pm=(1\pm \delta k)$, $B_\pm=((\nu-d)/q\mp q/2)$, in the rescaled units $m k_F/2 \hbar^2$ for $\Pi_{+-}$  the imaginary part reads:
\begin{equation}
\Im  \Pi_{+-}=\frac{-1}{4\pi q}
\begin{cases}
 A_+^2-B_+^2, & \text{in (a)}  \\
A_+^2-B_+^2-A_-^2+B_-^2, & \text{in (b) }\\
A_-^2-B_-^2, & \text{in (c) }\\
\end{cases}
\end{equation}
where the regions in the $(\nu,q)$ plane are defined by the following inequalities:  (a) ${\rm max}\{\nu_-^{(2)},\nu_+^{(2)} \}\le \nu\le \nu_+^{(1)}$ or $\nu_+^{(2)}\le\nu\le\nu_-^{(1)}$; 
(b) ${\rm max}\{\nu_-^{(1)},\nu_+^{(2)} \}\le\nu\le \nu_-^{(2)}$; and (c) $\nu_-^{(1)}\le\nu\le {\rm min}\{\nu_+^{(2)},\nu_-^{(2)} \}$.  
 Correspondingly, for the real part we have
\begin{equation}
\begin{split}\label{eq:super_re}
\Re \Pi_{+-} =
 \frac{1}{2\pi^2q} \left [A_+B_+-A_-B_-\right. \\\left.+\frac 1 2 ( A_+^2-B_+^2) \ln \left | \frac{A_++B_+}{A_+-B_+}\right|\right. \\\left.+\frac 1 2 ( A_-^2-B_-^2) \ln \left | \frac{A_-+B_-}{A_--B_-}\right|\right]
\end{split}
\end{equation}

The above results for the Lindhard function allow to obtain the  final result for the imaginary part of the spin-flip susceptibility
\begin{eqnarray}
\label{eq:mag_stoner}
\Im \chi_{+-}(\omega, q) = \frac{\Im \Pi_{+-}(\omega, q)}{(1+ g\Re \Pi_{+-}(\omega, q))^2 + g^2 \Im^2\Pi_{+-}} \nonumber \\+ A(q) \delta (\omega - \Omega_q),
\end{eqnarray}
this includes both the Stoner particle-hole continuum $ \Im \chi_{+-}^{Ston}(\omega, q)$ and the magnon contribution with weight $A(q)=(g^2 |\partial \Pi_{+-}/\partial \omega|_{\Omega_q})^{-1}$.

The RPA expression for the spin-flip susceptibility satisfies the sum rule \cite{Vignale:1}
\begin{equation}\label{eq:sum_rule}
-\frac{1}{\pi} \int_{-\infty}^{\infty}d\omega  \Im \chi_{+-}(\omega, q) =  \frac{(n_+-n_-)}{\hbar} .
\end{equation}
This provides an alternative way to fix the weight $A(q)$ of the collective magnon  mode with respect to the contribution of particle-hole excitations, according to the relation  $A(q)=(n_+-n_-)/\hbar + (1/\pi)\int_{-\infty}^{\infty}d\omega \Im \chi_{+-}^{Ston}(\omega, q)$. At increasing $q$ values the magnon strength decreases till it vanishes when the magnon mode enters the particle-hole continuum.
Nevertheless, as it is shown in Fig.\ref{fig1}, quite remarkably some trace of the magnon mode is left as a pronounced maximum in the particle-hole continuum. This feature
 could be used to infer the presence of the magnon mode even for wavevectors larger than those associated to the Stoner gap. 

{\it Spin-flip response for a trapped gas}

\begin{figure}
\includegraphics[width=0.9\linewidth]{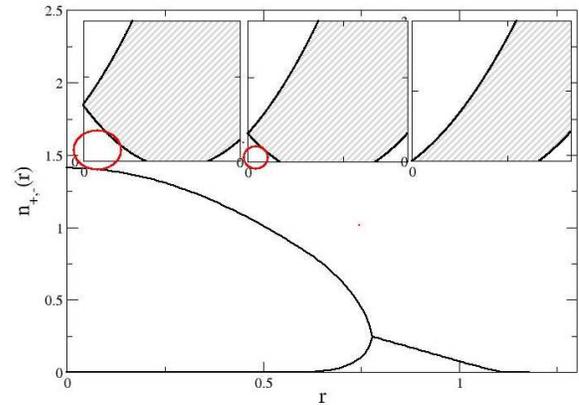}%
\caption{\label{fig2} (Color online) Main panel: Spatial density profiles $n_+(r)$ (top) and $n_-(r)$ (bottom), in units of the corresponding noninteracting densities at the trap center $n_\pm^{(0)}(0)$  as a function of the radial distance $r$ in units of $a_{ho}$ for  dimensionless interaction strength  $\lambda=k_F^0(0)a=2.5$, with $k_F^0(0)=(6 \pi^2n_\sigma^{(0)}(0) )^{1/3}$, and for a particle number $N=10^7$. Insets: the gray area indicates the Stoner particle-hole continuum in the $(\omega,q)$ plane as seen at various regions in the trap (center, middle, and sides from left to right) in a local-density picture, and the red circle indicates the region where magnon excitations are undamped at zero temperature.}
\end{figure}

We consider now the effect of inhomogeneity in the Fermi gas  induced by the presence of the external harmonic confinement  $V_{ext}(r)=\frac{1}{2}m\Omega_{tr}^2r^2$, taken for simplicity isotropic.
The equilibrium  density profiles $n_+(\B{r})$, $n_-(\B{r})$ of the two magnetization components  are found by minimizing the Thomas-Fermi energy functional \cite{Minguzzi:1,Sogo:1}
\begin{eqnarray}
E[n_\sigma(\B{r})] &=& \int d\B{r} \left \{\sum_\sigma \left[ \frac{3}{5} \alpha  n_\sigma(\B{r})^{5/3}  + \frac{1}{2}m\Omega_{tr}^2r^2 n_\sigma(\B{r}) \right. \right. \nonumber \\ &&\left. \left. -\mu_\sigma  n_\sigma(\B{r})\right] + gn_+(\B{r})n_-(\B{r})\right \}
\end{eqnarray}
where $\alpha = (6\pi^2)^{2/3}\hbar^2/2m$. The chemical potentials $\mu_\sigma$ are related to  particle numbers through the normalization conditions $\int d\B{r} n_\sigma(\B{r}) = N_\sigma$.  Note that  for simplicity the calculations have been performed at $\mu_+=\mu_-$, corresponding to fix the total number of particles $N$. An analysis beyond Thomas-Fermi approximation \cite{Dong10} would provide corrections at the tails of the density profiles but not alter significantly the resulting profiles. 

The dynamical spin-flip response of the trapped system  is obtained from the one of the homogeneous system through the local-density approximation (LDA) as an  average over the various trap regions,
\begin{equation}\label{eq:struc_LDA}
- g\Im_{+-}^{LDA} (\omega,\B{q} ) = \frac{-\int_{V_{tr}} d\B{r} g \Im \chi_{+-}(\omega, \B{q}, \B{r})}
{\int_{V_{tr}} d\B{r} },
\end{equation}
where $\Im \chi_{+-}(\omega, \B{q}, \B{r})$ is the homogeneous result (\ref{eq:mag_stoner}) which depends on $\B{r}$ through $n_+(\B{r}), n_-(\B{r})$, and $V_{tr}$ indicates the spatial volume where the particle densities are non vanishing. 
The local-density approximation is valid if the typical excitation frequency and wavevectors are larger than harmonic-oscillator frequency $\Omega_{tr}$ and the inverse size of the cloud.
Figure \ref{fig2} illustrates the particle density profiles and the Stoner particle-hole continuum corresponding to different  trap regions: while in the trap center the Stoner gap is maximal and magnon propagation is allowed, at the trap sides the gap closes and only the particle-hole continuum contributes to the dynamical response. 
 
\begin{figure}
\includegraphics[width=0.6\linewidth,angle=270]{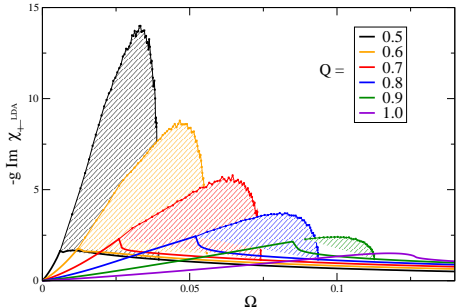}%
\caption{\label{fig3} (Color online) Imaginary part of the spin-flip susceptibility $-g$Im$\chi_{+-}(q,\omega)$ for a harmonically trapped Fermi gas as a function of the rescaled frequency $\Omega=m \omega/\hbar k_F^0(0)^2$ at various values of the transferred momentum $Q=q/k_F^0(0)$ for  dimensionless interaction strength  $\lambda=2.5$. The hetched areas indicate the magnon collective excitation contribution to the spectrum,  the solid lines indicate the contribution of the Stoner continuum which displays a maximum when the magnon merges onto it.}
\end{figure}

We illustrate in Fig.\ref{fig3} the total spin-flip response as a function of the frequency, for values of system parameters accessible in current experiments. 
Even in the presence of the external confinement we find that at small transferred wavevectors the dynamical response is dominated by the magnon contribution, emerging from the particle-hole continuum.  The latter acquires more importance at larger values of wavevectors. We expect this picture to hold at finite temperature, provided that it is smaller than the Stoner excitation gap $\Delta$.

{\it Conclusions}
 In this work we have considered the dynamical spin-flip response of a repulsive Fermi gas above the ferromagnetic transition, which is characterized by a magnon collective excitation mode. Within the random-phase approximation  we have provided an analytic expression for the the spin-flip response of the homogeneous two-component Fermi gas. For the experimentally relevant situation of an inhomogeneous gas, we have shown that even in the presence of an external confinement the dynamical response displays features of the magnon mode in a pronounced peak, thus displaying a clear indication of the ferromagnetic phase. 
These predictions are experimentally accessible via spin-dependent Bragg spectroscopy, which could be realized, similarly to the usual Bragg spectroscopy,  by a two-photon process inducing  transitions from $|\uparrow\rangle$ fermions to $|\downarrow\rangle$ fermions, transferring at the same time momentum $\hbar\B{q}$ and energy $\hbar \omega$ to the fluid. 

Refinements of our model include  the development of  a fully quantum description for the spectrum of the trapped interacting gas both at RPA level as done for the paramagnetic phase in \cite{Capuzzi01}, and beyond RPA as is done eg for attractive homogeneous  Fermi gases in \cite{Pieri:1}. A more accurate description of the experimental situation would require to include  in the dynamical description the effects of atom losses, and  the presence  of bound states as in \cite{Demler:1}.

\acknowledgments
We thank G. Conduit, J.N. Fuchs, W. Ketterle and L. Salasnich for discussions. We acknowledge support from the CNRS, the MIDAS project, the HANDY-Q project  and the PEPS-PTI project ``Quantum gases and condensed matter''.


\end{document}